\documentclass[preprint,aps,prl]{revtex4}

\usepackage{amsmath} 
\usepackage{graphicx}
\usepackage{bm}
\usepackage{amsfonts}

\begin{document}

\title{Characterizing dynamic length scales in glass-forming liquids}

\author{Elijah Flenner and Grzegorz Szamel}

\affiliation{Department of Chemistry, 
Colorado State University, Fort Collins, CO 80523
}
\maketitle
Recently, Kob \textit{et al.}\ \cite{Kob} reported non-monotonic temperature
dependence of a dynamic length scale, $\xi^{\mathrm{dyn}}$, in a model supercooled fluid. 
Specifically, they found that $\xi^{\mathrm{dyn}}$ peaks around a temperature at which deviations from mode-coupling-like
fits to the relaxation time become visible ($T=6.0$ for their model) and that there is a decrease of $\xi^{\mathrm{dyn}}$ 
below the mode-coupling crossover temperature ($T_c=5.2$ for their model). According to Ref. \cite{Kob}, 
the maximum of $\xi^{\mathrm{dyn}}$ signals a profound change in particles' motions. Its presence 
allows to identify the mode-coupling crossover without relying on any fitting procedure. 
It is consistent with a non-monotonic temperature dependence of finite size effects \cite{BBCKT}. 

To determine $\xi^{\mathrm{dyn}}$, Kob \textit{et al.}\ froze all particles in a semi-infinite space 
in an equilibrium configuration and analyzed the dynamics of the particles in the other half-space  
as a function of the distance from the boundary. This ``point-to-set'' method to determine a dynamic length scale   
differs from earlier calculations \cite{Sharon,FlennerL2010} where the spatial correlations of particles' dynamics were  
analyzed. By examining these correlations, we found that the more commonly used \cite{Sharon} 
dynamic correlation length $\xi_4$ increases monotonically with decreasing temperature in the range of temperatures
investigated by Kob \textit{et al}.

We simulated the binary harmonic sphere system of Kob \textit{et al.}\ using system sizes ranging 
from $N = 10,000$ to $100,000$ particles with up to $10^9$ time steps for the $100,000$ particle system at $T=5$. 
To determine $\xi_4$, we followed Refs. \cite{FlennerL2010,Flenner2011}. 
We defined a microscopic overlap function $w_n(t) = \Theta[a-|\mathbf{r}_n(t) - \mathbf{r}_n(0)|]$ 
where $\mathbf{r}_n(t)$ is the position of particle $n$ at a time $t$,
$\Theta(x)$ is the Heaviside step function, and $a=0.3\, \sigma$ ($\sigma$ is the diameter of the smaller particle). We calculated 
$S_4(q;t) = N^{-1} \left< \sum_{n,m} w_n(t) w_m(t) \exp[\mathbf{q} \cdot (\mathbf{r}_n(0)-\mathbf{r}_m(0)] \right>$,
which is the structure factor calculated using the initial positions of the particles that have moved less than a distance 
$a$ over a time $t$. Long range correlations of these particles 
are revealed by an increase of $S_4(q;t)$ at small $q$. To facilitate the fitting procedure 
used to determine $\xi_4(t)$, we independently calculated  
$\chi_4(t) = \lim_{q\to 0} S_4(q;t)$ utilizing 
the method proposed by Berthier \textit{et al.} \cite{Berthier} and described in detail in Ref. \cite{Flenner2011}.

\begin{figure}
\includegraphics[width=5in]{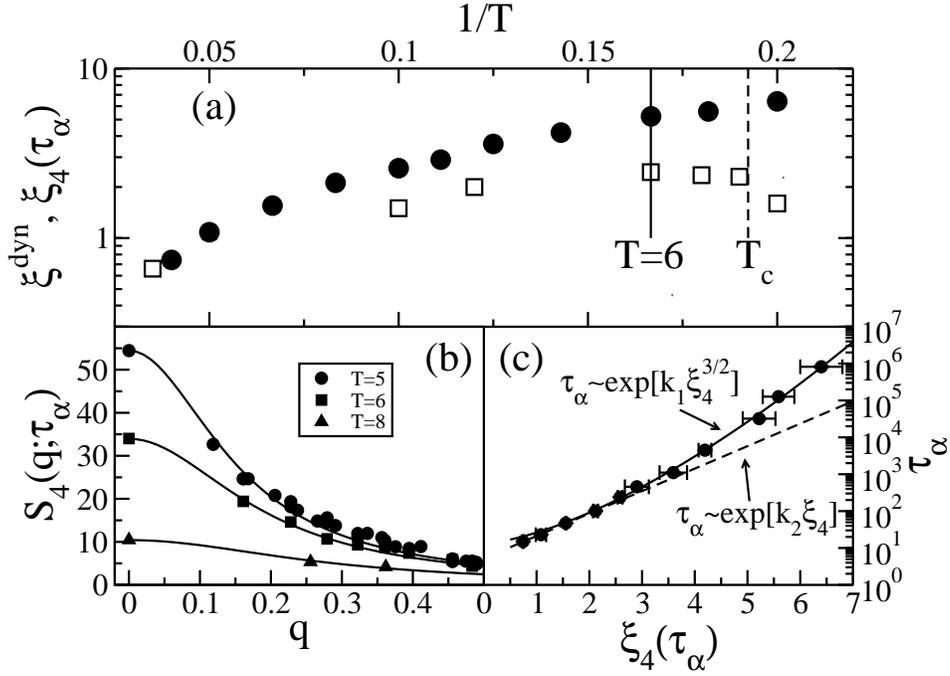}
\caption{\label{length}
\textbf{The temperature dependence of dynamic correlation lengths, the four-point
structure factor, and the relationship between the length and the relaxation 
time.}
(a) A comparison of the 
temperature dependence of $\xi_4(\tau_\alpha)$ (circles) calculated in this work and $\xi^{\mathrm{dyn}}$ (squares) 
from Fig. 3 of Ref. \cite{Kob}. (b) The
four-point structure factor $S_4(q;\tau_\alpha)$ (symbols) and 
the Ornstein-Zernicke fits, $\chi_4(\tau_\alpha)/[1+(\xi(\tau_\alpha) q)^2]$, 
used to determine $\xi_4(\tau_\alpha)$ (solid lines). (c) The relationship between 
$\tau_\alpha$ and $\xi_4(\tau_\alpha)$ showing a crossover from $\tau_\alpha \sim e^{k_1 \xi_4(\tau_\alpha)}$ 
to $\tau_\alpha \sim e^{k_2 \xi_4^{3/2}(\tau_\alpha)}$.
}
\end{figure}

In Fig.~\ref{length}(a) we compare $\xi_4(\tau_\alpha)$ (circles) and Kob \textit{et al.}'s $\xi^{\mathrm{dyn}}$ (squares). 
The relaxation time $\tau_\alpha$ is defined through $N^{-1}\left<\sum_n w_n(\tau_\alpha) \right> = e^{-1}$. 
In contrast to $\xi^{\mathrm{dyn}}$, 
the dynamic correlation length 
$\xi_4(\tau_\alpha)$ is monotonically increasing with decreasing temperature. This is also evident from a direct 
examination of $S_4(q;\tau_\alpha)$, Fig.~\ref{length}(b): it is readily apparent that $\xi_4(\tau_\alpha)$
at $T=5$ (circles) cannot be smaller than at $T=8$ (triangles), as was reported for $\xi^{\mathrm{dyn}}$. 
We shall emphasize that the finite wavevector data in Fig. 1(b) are 
ensemble-independent. As discussed in Ref. [6], the integrals of the 
four-point correlation functions, \textit{i.e.} the q=0 values in Fig.
1(b), are ensemble-dependent and this dependence is removed by adding
correction terms derived in Ref. [5].

We do, however, find evidence for a change of the particles' dynamics in the temperature range examined 
by Kob \textit{et al}, albeit at a slightly higher temperature, $T \approx 8$, than $T\approx 6$ suggested in Ref. \cite{Kob}. 
By examining the relationship between the self-diffusion coefficient, $\tau_\alpha$,
$\xi_4(\tau_\alpha)$, and $\chi_4(\tau_\alpha)$ we conclude that there is a crossover which manifests itself, \textit{inter alia}, 
in a change from 
$\tau_\alpha \sim \exp[k_2 \xi_4]$ to $\tau_\alpha \sim \exp[k_1 \xi_4^{3/2}]$, Fig.~\ref{length}(c).

We should emphasize the difference between $\xi^{\mathrm{dyn}}$ and $\xi_4$. The former length pertains to a non-linear
response of a fluid's dynamics to the pinning of one half of the particles whereas 
the latter characterizes the size of dynamic heterogeneities
in an unperturbed bulk fluid. While a length evaluated from a linear response function can be easily related to behavior in 
an unperturbed fluid, such a relation is much less clear for the non-linear response.   
Ref. \cite{Kob} and the present Correspondence highlight the importance of: (1) establishing 
a connection between non-linear response and dynamics in an unperturbed bulk fluid, (2) determining which length scale
is the most relevant one for the understanding of glassy dynamics in the bulk, and (3) 
extending simulations beyond the mode-coupling crossover.  

\section{Acknowledgements}

We gratefully acknowledge the support of NSF Grant CHE 0909676. This research utilized the CSU ISTeC Cray HPC System supported by NSF 
Grant CNS 0923386.

\end{document}